\input{psfig.sty}
\documentstyle[12pt]{article}
\newfont{\feff}{cmti10}
\begin{document}

\title{Self-sustained oscillations in  homogeneous shear flow. }

\author{ Victor Yakhot\\
Department of Aerospace and Mechanical Engineering\\
Boston University, 
Boston, MA 02215 }

\maketitle

${\bf Abstract}.$
\noindent Generation of the  
large-scale 
coherent vortical structurs  in homogeneous shear flow  couples  dynamical 
processes of  energy and enstrophy production. In the large rate 
of strain limit, 
the simple estimates of the contributions to
the energy and enstrophy equations result  in a dynamical system,  describing
experimentally and numerically observed 
self-sustained non-linear 
oscillations of energy and enstrophy. It is shown that 
the period  of these oscilaltions is independent upon the box size and 
the energy
and enstrophy fluctuations are strongly correlated.

\newpage
Due to its seeming simplicity,  the problem of  
homogeneous shear flow  has  widely been used as a benchmark   for
nu merical and experimental 
tests of various closures for turbulence modelling. All early closures were
based on the Kolmogorov ideas developed for statistically steady 
isotropic and homogeneous small-scale 
turbulence interacting with the non-universal large-scale flow-field. 
It became clear that to validate this physically appealing concept,  
one had  to verify and understand   the symmetries and other
statistical properties of the small-scale velocity fluctuations in the
real-life 
flows. This was the main focus  of the experimental studies of
homogenious shear flow [1]-[6]. The interest in this model flow is also
related to the recent 
 numerical investigations 
which revealed coherent structures resembling those, responsible for turbulence
production, in the wall- sheared flows
[7]-[13].  This system has also often  been   used for calibration of  
various constants  in semi-empirical turbulence models [15],[16].

The problem is formulated as follows:
consider a flow in a cube of a side $a$, so 
that $-a<x_{i}<a$. The velocity
field 

\begin{equation}
{\bf v}({\bf x},t)=U(y){\bf e_{1}}+{\bf u}({\bf x},t)
\end{equation}

\noindent with the imposed mean velocity $<\overline{{\bf v}}>=
U(y){\bf e}_{1}=Sy{\bf e}_{1}$ (the definition of the averaging operation will
be introduced below). The vorticity is defined then:

\begin{equation}
\Omega=-S{\bf e}_{3}+\omega
\end{equation}

The equations of motion for the fluctuating components of velocity and
vorticity 
are (density $\rho=1$):

\begin{equation}
\partial_{t}{\bf u}~ +~{\bf u}\cdot\nabla {\bf u}=-Sv{\bf e}_{1}-\nabla
p-U(y)\partial_{x}{\bf u}+\nu\nabla^{2}{\bf u},
\end{equation}

\begin{equation}
\partial_{t}\omega+{\bf u}\cdot\nabla\omega=\omega\cdot\nabla {\bf
u}-S(-\partial_{z}w{\bf e}_{3}+\partial_{x}w {\bf e}_{1}+\partial_{z}v{\bf e}_{2})-Sy\partial_{x}\omega
\end{equation}

\noindent and
\begin{equation}
\nabla\cdot{\bf u}=0
\end{equation}

\noindent  
The  $x$, $y$ and $z$-components 
 of velocity field are denoted hereafter as $u$, $v$ and $w$, respectively.  
Let us define the averaging
operations:

$$F=<\overline{F({\bf x},t)}>=\frac{1}{T}\frac{1}{V}\int_{0}^{T}\int_{V}d{\bf
x}dt
F({\bf x},t)$$

\noindent in the limit $V=a^{3}\rightarrow \infty; T\rightarrow \infty$. The
statistically steady state is assumed here. The spacial averaging is defined
as 

$$F(t)=\overline{F({\bf x},t)}=\frac{1}{V}\int_{V}d{\bf
x}
F({\bf x},t)$$

\noindent  The kinetic energy
equation is ($\nu\rightarrow 0$):

\begin{equation}
\partial_{t}{\cal K}+\frac{1}{2}\overline{ u_{i}\nabla_{i}
u_{j}^{2}}=-\tau_{uv}S-\nabla_{i}\overline
 {pu_{i}}-{\cal E}
\end{equation}

\noindent with $\tau_{vu}=\overline{uv}$. The contribution 
$\overline{U(y)\frac{\partial
u^{2}}{\partial x}}=0$ due to the symmetry of the problem.

Since in a homogeneous flow all spacial derivatives of the mean
properties are equal to zero, the modelling is reduced to investigation of the
time evolution of 
turbulent kinetic energy ${\cal K}=\overline{u_{i}^{2}/2}$ and  dissipation rate 
${\cal
E}=\nu\overline{(\partial_{i}u_{j})^{2}}$.

For small perturbations from isotropic and homogeneous state ($S\rightarrow 0$
and $\overline {pu}=\overline {pv}=0$), 
the typical turbulence models, based on 
the equilibrium ideas are [15], [16]:

\begin{equation}
\partial_{t}{\cal K}=-\tau_{ij}S_{ij}-{\cal E}
\end{equation}

\noindent and 

\begin{equation}
\partial_{t}{\cal E}=-C_{\epsilon 1}\tau_{ij}S_{ij}\frac{{\cal E}}{{\cal K}}-
C_{\epsilon 2}\frac{{\cal
E}^{2}}{{\cal K}}
\end{equation}

\noindent where the Reynolds stress 
$\tau_{ij}=\overline{v_{i}v_{j}}$. The coefficients
$C_{\epsilon i}=O(1)$. A simple expression, valid at the long times $t>{\cal
K}/{\cal E}$,  

\begin{equation}
\tau_{ij}\approx  -\nu_{T}S_{ij}
\end{equation}

\noindent
with  turbulent viscosity $\nu_{T}\propto {\cal K}^{2}/{\cal E}$,  closes the
set of equations (7)-(9) and defines the so called ${\cal K}-{\cal E}$ model,
widely used in engineering for modelling the not-too strongly sheared flows. 
 The unknown
magnitudes of the proportionality coefficients are typically determined in a following
way. Consider a flow  with $S=0$. The unknown coefficient $C_{\epsilon 2}$ can
be found from comparing the analytic solution of the simple equations (7)-(9)
with  experimental and numerical data. The same flow can also  be used to
test the results of analytic theories [15]-[16]. If $S\neq 0$, solution of
(7)-(9) is not easy and the coefficient $C_{\epsilon 1}$ can be found from
comparison with the data. The solution of equation (7)-(9) with the fixed  values
of the coefficients 
showed a close to 
exponential long-time growth of turbulent kinetic energy in a good agreement
with the outcome of direct numerical simulations [15]-[16].  
If the shear is imposed on a decaying isotropic 
turbulence at $t=0$,  
the observed [16]
initial, short- time decay of kinetic energy is readily explained by the fact
that turbulent viscosity 

$$\nu_{T}\approx \int_{0}^{t}\overline{v(0)v(\tau)}d\tau$$

\noindent which is small at short times.

Recent numerical experiments revealed  a much more complicated picture.
Driven
by a very strong shear (the criterion is derived below),  in the
long-time limit, 
 the
system developed a  limit cycle -like 
strong fluctuations of the total kinetic energy about the
mean value $<\overline{{\cal K}}>$ [7]-[13].
 The amplitude of these fluctuations was up to two-three times
that of  $<\overline{{\cal K}}>$.  Similar effect was observed by Borue et. al. [14] 
in a
three-dimensional Kolmogorov flow driven by a steady forcing 
${\bf f}=(0,0, cos(x))$. Elucidation of the origin of these oscillations is
the goal of this paper. 

The physical process  observed in both homogeneous shear and Kolmogorov flows
can be described in two steps: first,  the shear generates both kinetic energy
and  vortical structures leading to the access of the energy
production. Then, the structures become unstable and rapidly disappear with
the energy dissipation taking over. The process repeats itself. The
evolution  of kinetic energy and enstrophy fluctuations in 
3D Kolmogorov flow, conducted by Borue et al [14], revealed extremely strong
correlation: the sharp spikes in  the enstrophy and energy time-signals 
 were
almost simultanious with a slight time-lag, thus  suporting the importance of coherent
vortical structures in the process.

At the long times the numerical  homogeneous shear flow problem (1)-(3) 
has two very important
 features. We can see from the equation of motion that the flow, defined on a
 cube,
 cannot be
 periodic in space. Second, the integral scale ${\cal L}$ in this situation is not
 a dynamic
 variable which is a function of ${\cal K}$ and ${\cal E}$,  but prescribed by
 the box size, so that ${\cal L}\approx a$. This puts 
 strong constrains  on
 the modelling of various contributions to the equations  (1)-(3). 

Now, we would like to establish the main characteristic length-scales. The non-universal
velocity fluctuations belong to the range of scales $a\approx {\cal
L}<r<r_{c}$ 
with the cross-over scale $r_{c}\approx 
\sqrt{{\cal E}}/S^{\frac{3}{2}}$ are   dominated by  powerful 
anisortopic 
coherent structures (vortices). The univesal range,  populated by the more or
less isotropic excitations,  spreads over the interval $r_{c}<r<r_{d}\approx
(\nu^{3}/{\cal E})^{\frac{1}{4}}$. The Kolmogorov spectrum can be expected in 
the range with the
total energy of quasy-isotropic fluctuations 

\begin{equation}
q\approx \int^{r_{c}}_{r_{d}}E(k)dk\propto 
r_{c}^{\frac{2}{3}}- r_{d}^{\frac{2}{3}}>0
\end{equation}

We can see that the inertial range shrinks to zero when the strain rate
becomes large. This fact, noticed in Ref. [13], defines the strong shear
regime. 
In strongly anisortopic flow, the
simple expression (9), is invalid.

First , let us consider the equation for $\overline{\omega^{2}}$:

\begin{equation}
\frac{1}{2}\partial_{t}\overline{\omega^{2}}+\frac{1}{2}
\partial_{i}\overline{u_{i}\omega^{2}}=
\overline{\omega\cdot\omega\cdot\nabla{\bf
u}}-
S\overline{(-\omega_{z}\partial_{z}w+\omega_{1}\partial_{x}w +\omega_{2}\partial_{z}v )}                           -
\nu\overline{(\partial_{i}\omega_{j})^{2}}
\end{equation}

\noindent Due to powerful, shear-generated  coherent  vortical structures, 
$\overline{\omega_{y}\omega_{x}}=O(\overline{\omega^{2}})$ and the
contribution involving longitudinal derivative $\partial_{z} w$ can be neglected. 
The simple dimensional considerations lead
to:

$$ \overline{\omega\cdot\omega\cdot\nabla{\bf u}}=O(u_{rms}\omega^{2}/{\cal L});~~
S\overline{(\omega_{x}\partial_{x}w+\omega_{y}\partial_{z}v)}=O(S\overline{\omega^{2}})$$

\noindent and denoting $A(t)=\overline{\omega^{2}}/S^{2}$ we have:

\begin{equation}
\partial_{t}A(t)=\gamma(\frac{a}{{\cal L}S}\sqrt{{\cal K}}-\alpha)SA(t)
\end{equation}

\noindent with all coefficients $\gamma$, $a$ and $\alpha=O(1)$.

 To
estimate $\tau_{uv}$, we observe  that in the limit of interest 
(see below) the only
relevant time scale is $\tau_{c}\approx 1/S$. From the equation (3)
we have an estimate for the stress:

\begin{equation}
\tau_{uv}=\overline{u(t)v(t)}\approx \overline{u(t)\int_{t-\tau_{c}}^{t} d\lambda 
(-{\bf u}(\lambda)\cdot \nabla v(\lambda)-\nabla_{y}p(\lambda))}
\end{equation}

\noindent where the ``initial condition $\tau_{uv}(t-\tau_{c})$ was
neglected for simplicity (see below).  The dimensional estimate gives:

\begin{equation}
\tau_{uv}=B(t)\frac{{\cal K}^{\frac{3}{2}}}{S{\cal L}}
\end{equation}

\noindent where $B(t)$ is an ``anisotropy factor'' or ``order parameter'' 
characterizing the varying
in time strength
of the coherent vortical structures.  The appearence of this factor is natural
(see below)
since 
in an  isotropic,  non-sheared flow lacking coherent
vortices $B(t)=const=0$, while in the strongly sheared 
flow these structures contribute to the
energy production. 
The dissipation rate in shear flows is estimated as 
${\cal E}=b{\cal K}^{\frac{3}{2}}/{\cal L}$ with the coefficient $b\approx 1$
leading to a model equation:

\begin{equation}
{\cal K}_{t}(t)\approx (B(t)-b)\frac{{\cal
K}^{\frac{3}{2}}}{{\cal L}}
\end{equation}

\noindent or, introducing $y=\sqrt{{\cal K}}$:

\begin{equation}
2\partial_{t}y(t)=
\frac{B(t)-b}{{\cal L}}y^{2}
\end{equation}

By the virtue of (2),
the mean value of vorticity in homogeneous shear flow is 
$<\overline{\Omega}>=S$. Thus, the natural measure of the 
strength  of the strongly anisotropic fluctuating coherent vortical structures  is 
the ratio $\overline{\omega}^{2}/S^{2}$. Based on these considerations , 
we set $A(t)=B(t)$. This result can be derived in the third-order of the
iteration procedure of the expression (13). Indeed, inserting an unknown
initial condition into (13), one can use  $\tau_{uv}(t-\tau_{c})=\tau_{0}$
as a zero order solution. Then, after simple resummation (neglecting  the first
-order contributions)  we obtain:

$$\tau_{uv}(t)=\tau_{0}+\int_{t-\tau_{c}}^{t}\overline{u(t){\bf u(\lambda)\cdot \nabla}
v(\lambda)}+
\int\int\int \overline{ u(t){\bf u(\lambda)\cdot\nabla u(\lambda')\cdot\nabla
u(\lambda '')\cdot\nabla }v(\lambda'')}d\lambda d\lambda'
d\lambda'' + \cdot\cdot\cdot $$

This expression immediately gives (14) with $B(\omega)\approx
\frac{\omega^{2}}{S^{2}}$,
provided the derivatives $\partial_{i}u_{j}\approx \omega$. 
This approximate derivation is given here to demonstrate the mechanism of
vorticity 
appearence  in the expression for the reynolds stress $\tau_{uv}$. It will
become clear below that the power of vorticity in the expression
$B(\omega)\propto \omega^{n}$ is unimportant.

The fact that the anisotropic ordered structures can influence the magnitude and even
sign of the energy fluxes is known  for a long time. 
The rigorous linear 
stability analysis developed 
in [17]-[18], showed that even in  three-dimensional flows the 
strongly anisotropic structures (basic flows),
are capable of reversing the sign of the
energy flux,  due to the  ``negative
viscosity'' effects  and lead to substantial growth 
of a  small large-scale perturbation. In the opposite limit  of the 
 isotropic basic flows, the theory showed 
generation of positive effective 
viscosity and  acceleration of  the energy dissipation. 
 These features are
incorporated  in the model (12),(15), (16): 
indeed, we see that if $A(t)>b$, the energy
grows, while,  when $A<b$, it decays.  The model equation (12) includes the
well-known process of the vortex break -down: when $v^{2}>>\omega^{2}{\cal L}^{2}$,
the instability leads to the vortex disappearence. 

\noindent Defining dimensionless variables $Z=\frac{y}{S{\cal L}}$, 
$Z_{0}=\frac{\alpha}{a}$ and $T=St$ gives:

\begin{equation}
2\partial_{T}Z(T)=a(A(T)-b)Z^{2}(T)
\end{equation}

\noindent and

\begin{equation}
\partial_{T} A(T)=-\gamma(Z(T)-Z_{0})A(T)
\end{equation}

\noindent where 
 $Z_{0}>0=const$, related to the mean amplitude of $Z(t)$. 

The equations (17),(18) have a steady-state solution 
$Z(T)=Z_{0}$  and $A(T)=b$. A simple linear stability analysis 
shows periodic solution when the amplitude of the perturbation is very 
small.  
The numerical solutions  of quations (17),(18), 
presented on Figs. 1-6, revealed 
 strong non-linear oscillations.  All calculations were perforemed with
$Mathematica^{TM}$.
In a wide range of parameter variation, the
system generates non-linear oscillations with the shape 
 depending upon initial values 
 $Z(0)$ and $A(0)$. For a given set of parameters the frequency of
oscillations is proportional to the strain rate $S$. 

For the initial values of $Z(0)$ and $A(0)\approx 1$,
the solution shows  reasonably smooth oscillations with $Z$ and $A$ being
somewhat out of phase  (see Figs. 1,2). 
 The result supports a  general physical picture of
the anisotropy $A(t)$ (order parameter) and energy growing  (decaying) 
together with
some time-lag. The energy fluctuations are by a factor 2-3 larger than
$S^{2}{\cal L}^{2}$. When the initial energy was doubled to $Z(0)=3.$, the
oscilaltions became much less symmetric with the steeper energy grows
(Figs. 3,4). The crucial role of the ``order parameter'' $A(0)$ is
demonstrated on Figs. 5,6 corresponding to $Z(0)=2$ and $A(0)=0.1$. We can see
the
the formation steep shock-like structures, somewhat resembling
turbulence-production bursts. 

 In the range of large
$Z(0)$ and very small $A(0)<<1$, the solution blows up, 
 indicating the unphysicality of these initial
conditions corresponding to the large energy fluctuations ($u_{rms}(0)>>S{\cal
L}$) and small anisortopy (order) parameter ($\overline{\omega^{2}}<<S^{2}$). 

To discuss the above results, let us 
look at  this work from a somewhat different angle.
The Kolmogorov relation $S_{3,0}(r)=\overline{(u(x+r)-u(x))^{3}}\propto r$, 
is a statement about constantcy of the  energy
flux for inertial range wave numbers  $k\approx 1/r>>1/{\cal L}$ of 
isotropic and homogeneous turbulence. As $r\rightarrow {\cal
L}$, the structure function $S_{3,0}(r)\rightarrow 0$. 
In strongly anisotropic flows with the
integral scalle ${\cal L}\approx a$, this is not so: depending on the 
spacial distribution of  velocity
(vorticity),   the moment 
$S_{3}({\cal L})\neq 0$. If vorticity (enstrophy) 
is an ``order'' parameter, characterizing deviations  from isotropy, then 
$S_{3}({\cal L})\approx B(\omega){\cal K}^{\frac{3}{2}}/{\cal L}$  where
$B(\omega)\rightarrow 0$  when the strength of the structures
diminishes. This qualitative statement 
is supported by the well-known fact that the
velocity field, generated by the vortex $v(r)\propto {\Gamma}\phi(r)$, where
the circulation 
$\Gamma=O(\omega {\cal L}^{2})$. 
Combined with the equation for the enstrophy, the two relations (17),(18)
form a  dynamical system leading to strong fluctuations of both energy
and enstrophy. The shape of the function $B(\omega)$ does not seem to
influence the qualitative aspects of the process: the model
(17),(18) is invariant  under transformation $A(t)\rightarrow A^{n}$ with a
simple rescaling of time.

All this is valid when  ${\cal L}\approx a$. If this is not so, the magnitude
of the fluctuations must substantially decrease. Indeed, if  $a>>{\cal L}$,
then we are dealing with $N=(\frac{a}{{\cal L}})^{d}$ independent
systems. Here $d$ is the force dimensionalty.
Since the phases are crucially important , we expect the amplitude of
the fluctuations to decrease as $1/\sqrt{N}$. This can easily  be tested on an
example of 3D Kolmogorov flow in a box with the side $a$ 
driven by the force ${\bf f}=(0,0,cos
(\frac{x}{{\cal L}}))$ by varying the forcing scale. 

To conclude: based on the equations of motion and some physical 
considerations, we
propose a dynamic model,  coupling vorticity (enstrophy) and energy 
fluctuations in a
homogeneous shear flow. This model generates strongly correlated
self-sustained oscillations of both enstrophy and energy similar to those
observed in eperiments and  direct numerical simulations.
The calculated time
-lag is similar to that  observed in a numerical study of 3D Kolomogorov flow
by Borue et al [14]. 

It is not yet clear if,  properly parametrized, this simple model can mimic
turbulent bursts which are at the core of the energy production 
in turbulent wall
flows. In case of a positive answer, the model of this kind can serve as a
boundary condition (``wall function'') 
for turbulence simulations,  neglecting  the detailed
consideration of dynamics of the viscous sublayer.  The achieved computational 
economy makes this aspect of the work worth persuing.

\noindent {\bf references}\\
\noindent 1. S.  Tavoularis and S. Corrsin,  J. Fluid Mech., {\bf 104}, 331,
 (1981)\\
\noindent 2. S. Tavoularis and S. Corrsin,  J. Fluid Mech., {\bf 104}, 349, 
(1981)\\
\noindent 3. F.H. Champain,  V.G. Harris,  and S. Corrsin  J. Fluid Mech., {\bf 41}, 81,(1970)\\
\noindent 4. Rose,  J. Fluid Mech., {\bf 25}, 97,(1966)\\
\noindent 5. S. Garg and Z. Wargaft, Phys. Fluids, {\bf 10}, 662, (1998)\\
\noindent 6. X. Shen and Z. Wargaft,  Phys. Fluids, {\bf 12}, 2976, (2000)\\
\noindent 7. W. Rogers and P. Moin, J. Fluid. Mech., {\bf 176}, 33, (1987)\\
\noindent  8. C.  Lee, J. Kim and P. Moin, J. Fluid. Mech., {\bf 216}, (1990)\\
\noindent 9. S. Kida and M. Tanaka, J. Fluid. Mech., {\bf 274}, (1994)\\
\noindent 10. K. Nomura, in  IUTAM Symposium on Geometry and Statistics of
turbulence, Kluwer Acad. Publishers, Boston, (2002)\\
\noindent 11. A. Pumir, Phys. Fluids, {\bf 8}, 3112,  (1996)\\
\noindent 12. A. Pumir and B. Shraiman (1995), Phys.Rev.Lett., {\bf 75}, 3114,(1995)\\ 
\noindent 13. P. Gualtieri, C.M.Casciola, R. Benzi, G. Amati and R.Piva,
arXiv:nlin.CD/0011040v2, 27Nov(2000)\\
\noindent 14. V. Borue and S. Orszag, J. Fluid. Mech.,.... (1996)\\
\noindent 15. B. Launder and D.B. Spalding, Com. Mech. in
Appl. Mech. Eng,. {\bf 3}, 269 (1974)\\ 
\noindent 16.  V. Yakhot, S. Orszag, S. Thangam, T. Gatsky and C. Speciale,
Phys. Fluids A, {\bf 4}, 1510 (1992)\\
\noindent 17. V.Yakhot and G. Sivashinsky, Phys. Rev. A35, 815  (1987)\\
\noindent 18. V. Yakhot and R. Pelz, Phys. Fluids. {\bf 30}, 1272  (1987)\\

..................
\\

\newpage

\begin{figure}[h]
\centerline{\psfig{file=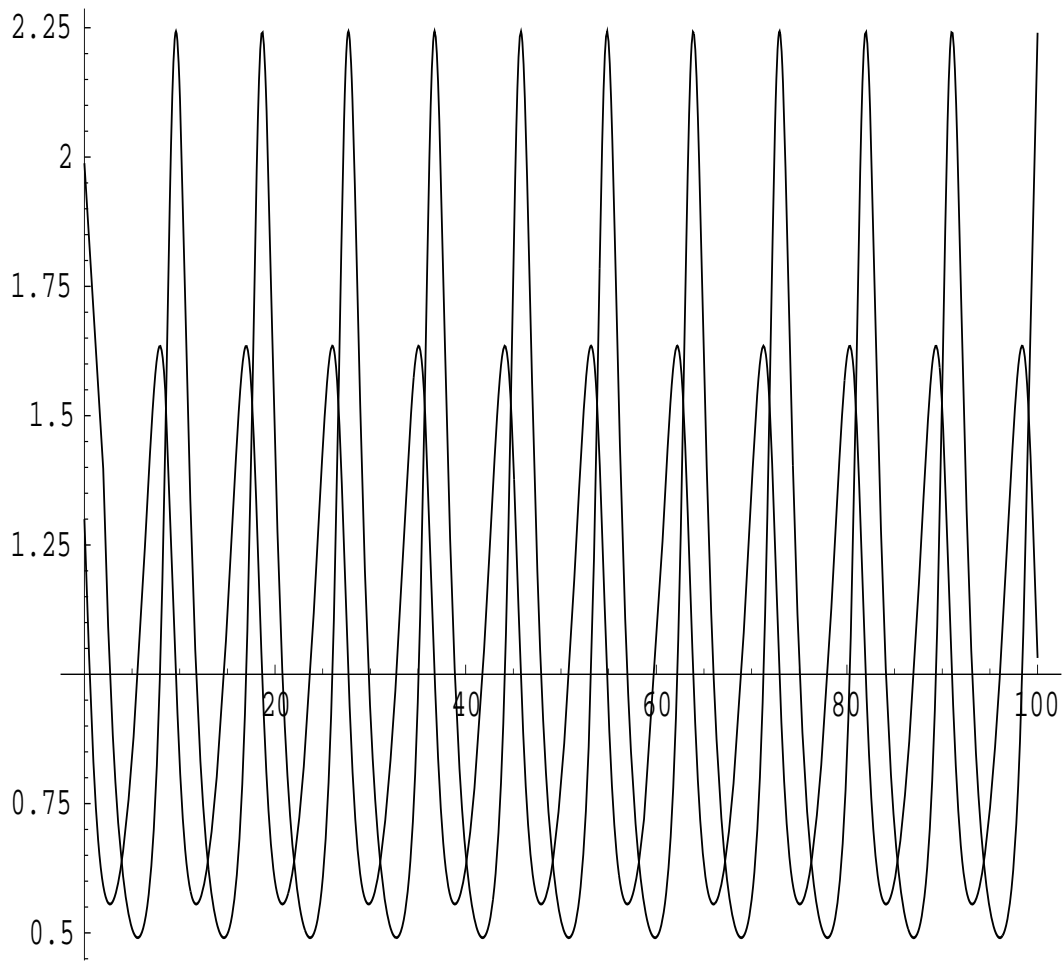,width=5.5in,height=5.0in}} 
\caption{Time -evolution of $Z^{2}(T)\propto {\cal K}(T)$ (higher amplitude
curve) and $A(T)$ vs $T$.  
$Z(0)=1.41$,~~$ A(0)=1.3$} 
\end{figure}

\begin{figure}[h]
\centerline{\psfig{file=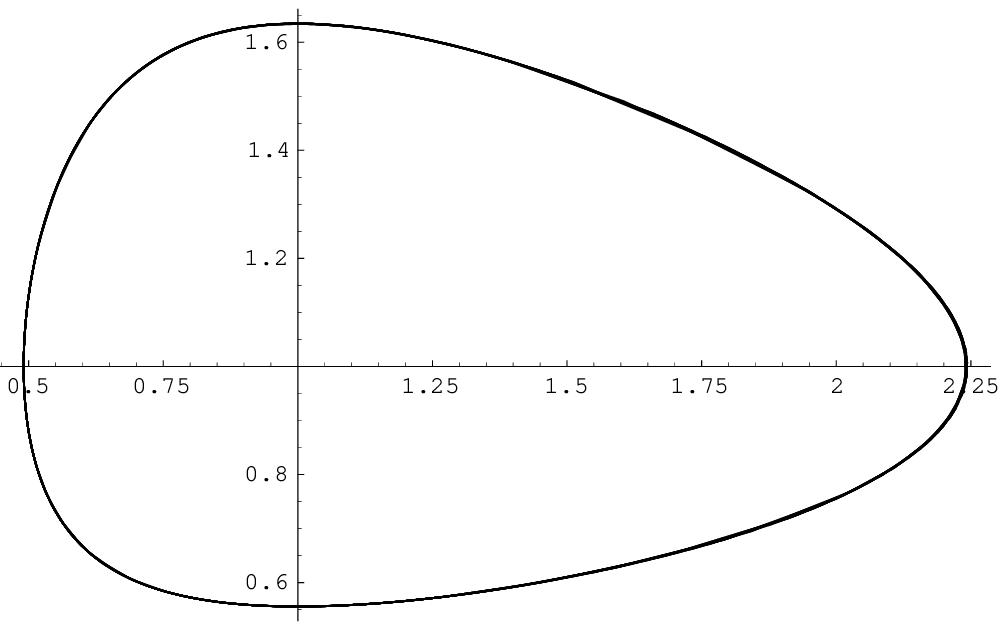,width=5.5in,height=5.0in}}
\caption{Parametric plot $Z^{2}(T)$ (horizontal) vs. $A(T)$}  
\end{figure}

\begin{figure}[h]
\centerline{\psfig{file=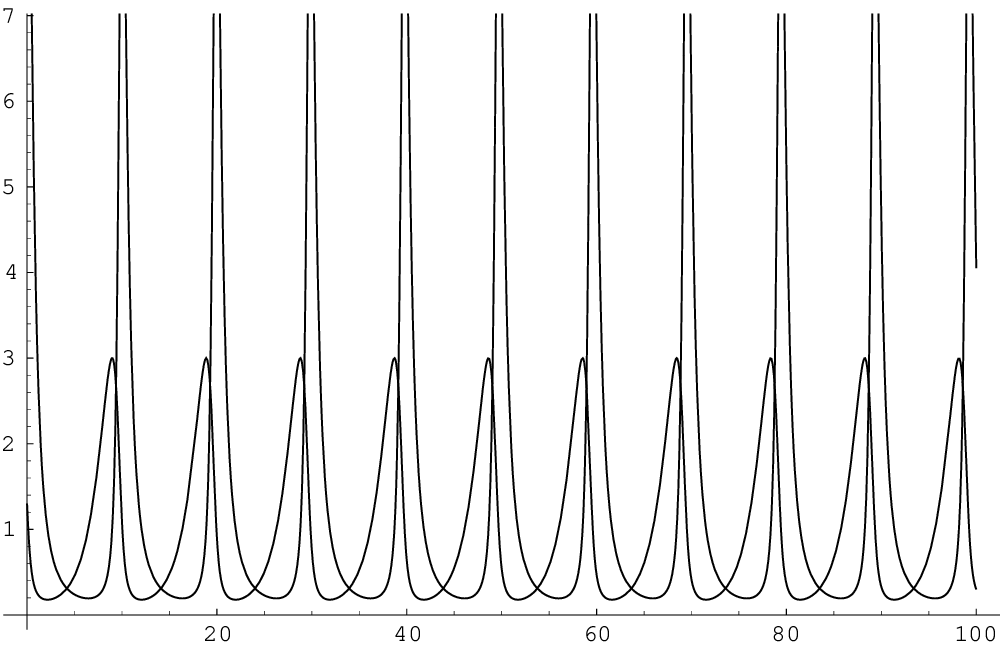,width=5.5in,height=5.0in}} 
\caption{Time -evolution of $Z^{2}(T)\propto {\cal K}(T)$ (higher amplitude
curve) and $A(T)$ vs $T$.  
$Z_{0}=3$; $A(0)=1.3$}  
\end{figure}

\begin{figure}[2^{b}]
\centerline{\psfig{file=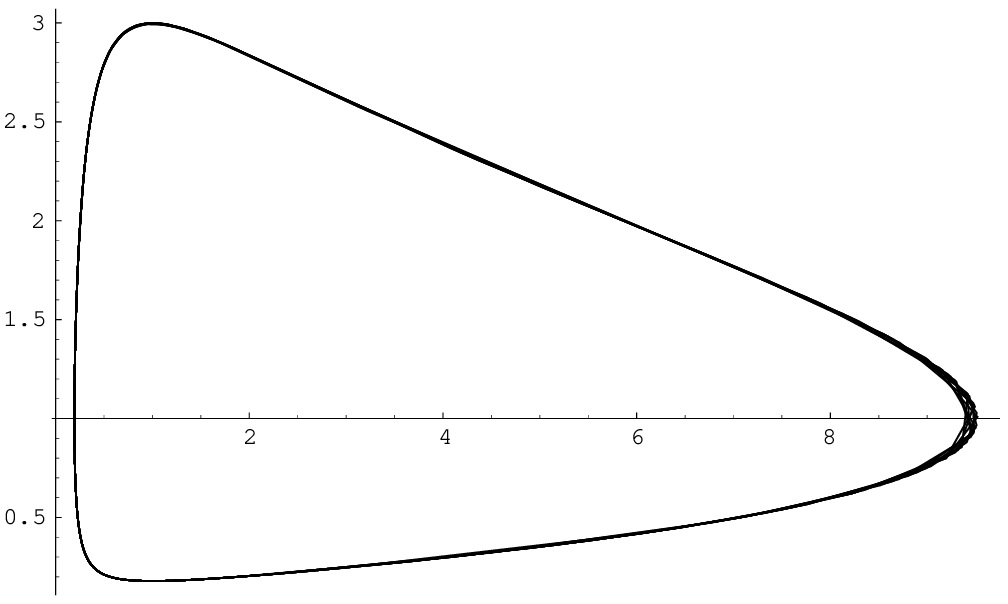,width=5.5in,height=5.0in}} 
\caption{Parametric plot $Z^{2}(T)$ (horizontal) vs $A(T)$} 
\end{figure}

\begin{figure}[h]
\centerline{\psfig{file=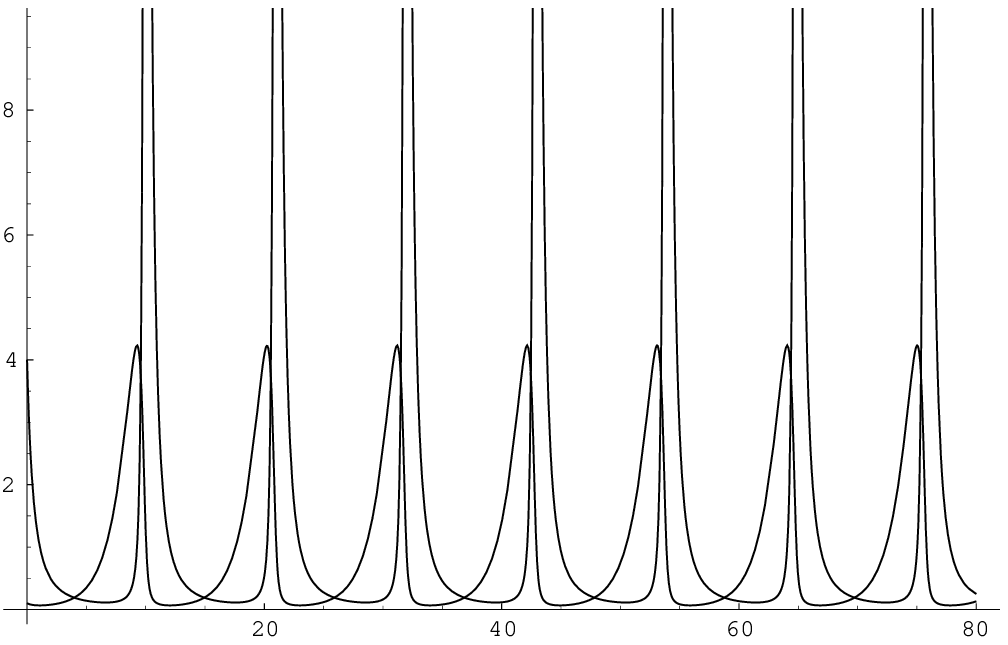,width=5.5in,height=5.0in}} 
\caption{Time -evolution of $Z^{2}(T)\propto {\cal K}(T)$ (higher amplitude
curve) and $A(T)$ vs $T$.  
$Z_{0}=2$; $A(0)=0.1$}  
\end{figure}

\begin{figure}[h]
\centerline{\psfig{file=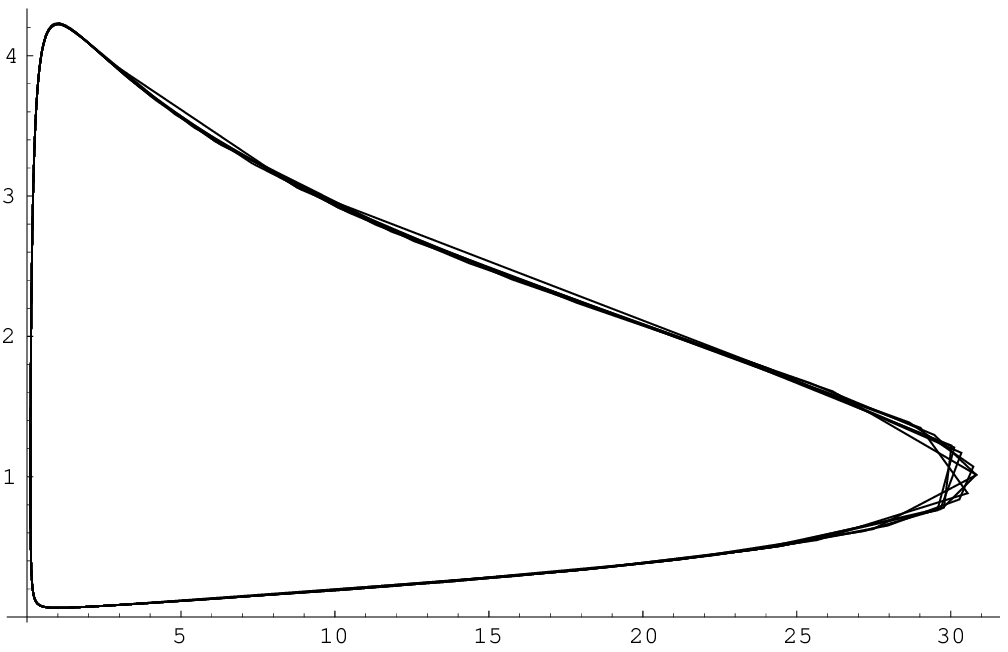,width=5.5in,height=5.0in}} 
\caption{Parametric plot $Z^{2}(T)$ (horizontal) vs $A(T)$} 
\end{figure}

\end{document}